\documentclass{article}

\pdfoutput=1

\usepackage{PRIMEarxiv}
\usepackage{amsthm,amssymb, amsmath}

\usepackage[utf8]{inputenc} 
\usepackage[T1]{fontenc}    
\usepackage{hyperref}       
\usepackage{url}            
\usepackage{booktabs}       
\usepackage{amsfonts}       
\usepackage{nicefrac}       
\usepackage{microtype}      
\usepackage{lipsum}
\usepackage{fancyhdr}       
\usepackage{graphicx}       
\graphicspath{{media/}}     

\title{Time-domain Hong-Ou-Mandel interference \\ of quasi-thermal fields and its application \\ in linear optical circuit characterization
}

\author{
  Anna Romanova\textsuperscript{1,*}, Konstantin Katamadze\textsuperscript{1,2}, Grant Avosopiants\textsuperscript{1}, Leon Biguaa\textsuperscript{1}, \\ \textbf{Nikolay Skryabin\textsuperscript{1}, Alexander Kalinkin\textsuperscript{1}, Ivan Dyakonov\textsuperscript{1}, Yurii Bogdanov\textsuperscript{2}, Sergei Kulik\textsuperscript{1}}\\
  \textsuperscript{1}Quantum Technology Centre, Faculty of Physics, Lomonosov Moscow State University, Moscow, Russia\\
   \textsuperscript{2}Valiev Institute of Physics and Technology, Russian Academy of Sciences, Moscow, Russia \\
   \textsuperscript{*}romanova.phys@gmail.com \\
}

\begin{document}
\maketitle

\begin{abstract}
We study temporal correlations of interfering quasi-thermal fields, obtained by scattering laser radiation on a rotating ground glass disk. We show that the Doppler effect causes oscillations in temporal cross-correlation function. Furthermore, we propose how to use Hong-Ou-Mandel interference of quasi-thermal fields in the time domain to characterize linear optical circuits.
\end{abstract}


The history of correlation measurements originates from the moment when Hanbury Brown and Twiss proposed the new type of interferometer
based on correlations between signals received by two independent detectors, which was robust to atmosphere fluctuations.  
%
%
Hanbury Brown and Twiss's research gave scientists a tool for studying photon correlations and accelerated the progress of quantum optics.
In particular, the Hong-Ou-Mandel (HOM) effect \cite{Hong1987}, which is based on correlation measurements of photon pairs, allows to measure time delays with high precision.
It also helps to distinguish quantum states of light from classical ones. Many experiments originally carried out with entangled photons were repeated with thermal light as well to check whether quantum correlations are necessary for observing the effect
~\cite{Liu2013, Gatti2004, Bobrov2013, Classen2016, Katamadze2018}. Such experiments actually use quasi-thermal sources allowing to control correlation characteristics of radiation. 
For example, they are created by scattering of laser radiation on a rotating ground glass disk
(RGGD)~\cite{Martienssen1964}.
The dependences of spatial coherence on the laser mode, and on the size of heterogeneities were investigated for RGGD~\cite{Asakura1970}. An analogue of HOM-interference of two uncorrelated thermal fields has been studied in spatial degrees of freedom~\cite{Chekhova1996, Liu2013}. However, the time dependence of the intensity correlation function for two interfering thermal fields has not been investigated in detail so far.

Nowadays, quantum technologies and particularly quantum computation are beginning to play an increasing role. One of the most promising platforms for creating quantum computer is optical realization. In particular, quantum advantage has recently been demonstrated using a bulk multiport interferometer~\cite{Zhong2020}. It is clear, however, that bulk optics can be used for proof-of-principle experiments only and miniaturization to integrated optical circuits~\cite{OBrien2009}, or chips, is necessary for large-scale quantum computation. Scalable quantum computing needs gate fidelity $>0.99$ to implement quantum error correction~\cite{Fowler2012}. Achieving such an accuracy requires good control over optical circuits, and consequently one needs to characterize them with high precision.

The process implemented by a linear optical circuit is represented as a $N \times N$ transfer matrix $\hat U$ with complex elements, which connects input $E^{\text{(in)}}$ and output $E^{\text{(out)}}$ field complex amplitudes:
\begin{equation} 
    E_i^{\text{(out)}} = \sum\limits_{j = 1}^N {{U_{ij}}E_j^{\text{(in)}}},
\end{equation}
where $ i, j = 1, 2,.., N $ and $N$ is a number of input and output channels.

Existing methods of characterization~\cite{Rahimi-Keshari2013, Heilmann2015, Peruzzo2011, Laing2012} do not allow reconstruction of all the elements of $\hat U$ up to their phases. At the same time, all the matrix element phase (MEP) values are not necessary to obtain, since in many applications the circuit inputs are fed with factorized Fock states and the measurement results are independent of the input and output phases~\cite{Laing2012}. Therefore, we will assume that any transfer matrix $\hat U$ can be reduced to an equivalent matrix $\hat M$ by a selection of diagonal matrices $\hat D_1$ and $\hat D_2$ that nullify the MEPs of the first row and the first column: $\hat M = D_1 \hat U D_2$. Thus, $N$-channel linear network can be characterized by $N^2$ absolute values $\{M_{ij}\}$ of matrix elements and $(N-1)^2$ non-trivial MEPs.

The problem of reconstructing parameters of a transfer matrix of a multichannel linear optical circuit is reduced to reconstruction of its $2 \times 2$ submatrices~\cite{Laing2012}, which are generally non-unitary and have only one non-zero MEP. Absolute values of matrix elements are simply square roots of intensity transmission coefficients and are reconstructed trivially. MEP reconstruction is more complicated, so this is the problem discussed below.

The basic methods of characterization are the algorithm based on coherent states~\cite{Rahimi-Keshari2013, Heilmann2015} and the routine based on biphoton states~\cite{Peruzzo2011, Laing2012}. The first one is easy to perform, but unstable to fluctuations of relative phases of input states. Such fluctuations are inevitable if radiation is supplied to an optical chip through optical fibers. Although the use of Lissajou figures, as suggested in the work~\cite{Heilmann2015}, solves this problem, it becomes difficult to account for the acquisition electronics time jitter. The second one, essentially based on HOM interference, is free from an influence of input phase fluctuations, but requires to accumulate statistics of photocounts. Therefore sources with low efficiency lead to a long data acquisition time.

In our previous work~\cite{Katamadze2021}, we showed how interfering thermal fields can be used to characterize linear optical circuits. Like the Hanbury Brown -- Twiss stellar interferometry, this technique is based on intensity correlation measurements and is therefore robust to input phase fluctuations. In addition, it does not require a long acquisition time because it uses bright classical thermal fields. The efficiency of the method was numerically demonstrated by simulating measurements of the intensities $I_1(t)$ and $I_2(t)$ at two circuit outputs and calculating the value of a normalized second-order correlation function:
\begin{equation}
    g^{(2)}(\tau) = \frac{\langle I_1(t) I_2(t+\tau)\rangle} {\langle I_1(t) \rangle \langle I_2(t)\rangle}.
\end{equation}
We conducted the reconstruction of transfer matrix at $\tau = 0$, because the value $g^{(2)}(\tau=0)$ is directly dependent on the MEP. However, during experimental verification we encountered oscillations of temporal correlation function $g^{(2)}(\tau)$. Below we introduce the theoretical description of this effect and show how it benefits the linear optical circuit characterization. Then we conduct a series of experiments to reconstruct unknown MEPs of 2-mode processes and verify the experimental results by the reference method~\cite{Rahimi-Keshari2013, Heilmann2015}.

\begin{figure}[ht]
\centering\includegraphics[width=0.8\linewidth]{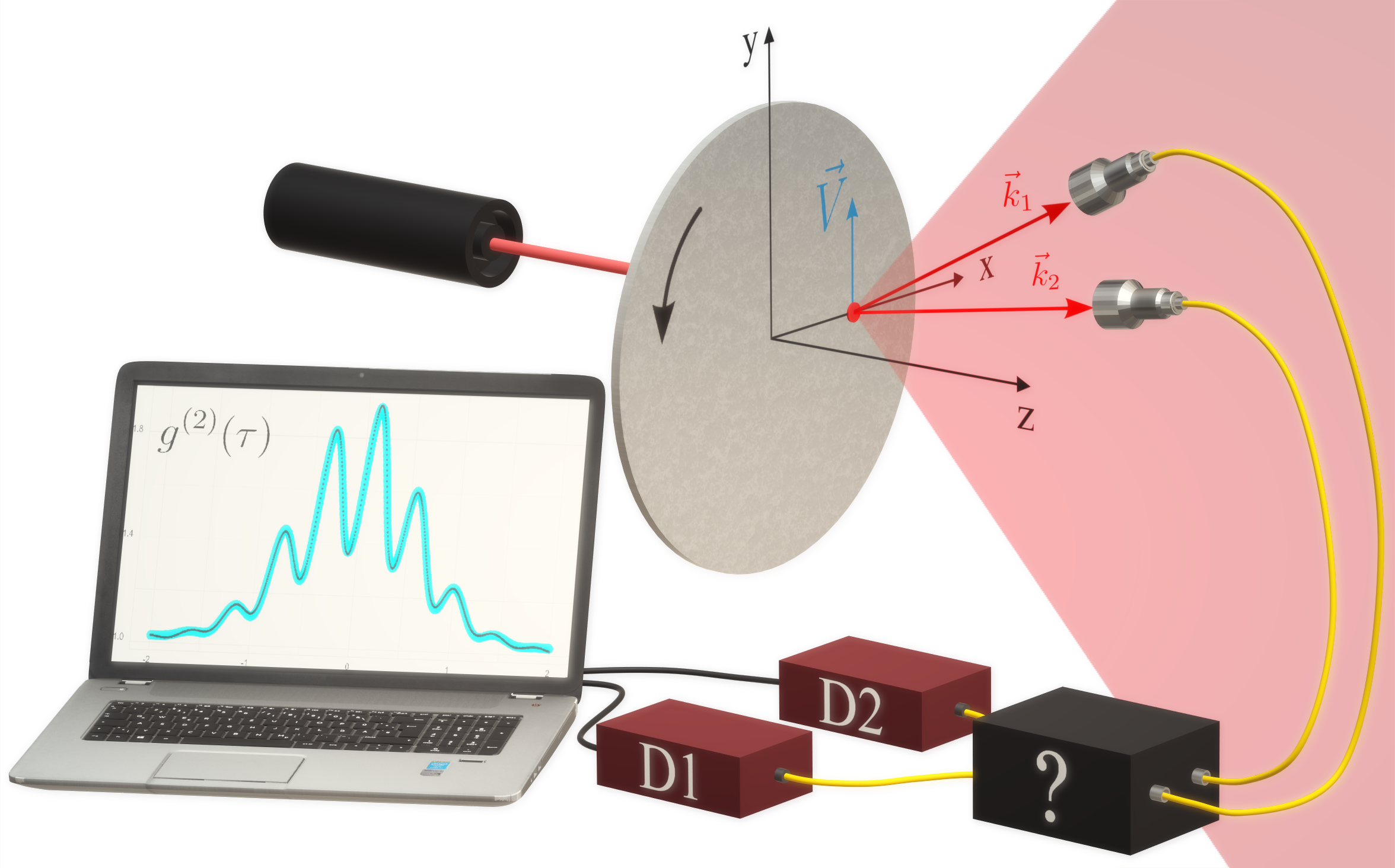}
\caption{General scheme for preparing thermal fields and black box network characterization. Laser radiation is scattered on the rotating ground glass disk. $\vec{V}$ is the speed of the disk at the waist point, $\vec{k_1}$ and $\vec{k_2}$ are the wave vectors in the direction of radiation collection. Two uncorrelated thermal fields are directed to the black box inputs. Output intensities are measured by analog detectors D1 and D2 and and detector output signals are digitized. The black box MEP is reconstructed based on the calculated values of intensity correlation function $g^{(2)}(\tau)$.}
\label{fig:disk}
\end{figure}
  
Let us derive an expression of the correlation function of the quasi-thermal field obtained by focusing a gaussian laser beam with the waist $w$ on the RGGD. The beam and disk centers are on the same level, so the disk velocity $\vec{V}$ at the beam center is directed vertically (See Fig.~\ref{fig:disk}). 
We consider fields scattered at an angle to the disk normal with the wave vectors $\vec{k}_{1,2}$. Following the formalism set forth in~\cite{Crosignani1971} (see details in Supplement) we obtain the degree of first-order coherence for each field:
\begin{equation}\label{eq:g_1}
    g^{(1)}_{1,2}(\tau) \varpropto \exp\left[-\cfrac{\tau^2 }{4 \sigma^2}\right]\exp \left[\text{i}(\omega+\Delta_{1,2})\tau \right] .
\end{equation}
Here, the first factor is the standard Gaussian envelope corresponding to the correlation properties of a thermal state, with the correlation time $\sigma=w/V$, which corresponds to the time of passage of the disk point through the beam waist. The second factor is sinusoidal oscillation, where $\omega$ is the frequency of the laser radiation and $\Delta_{1,2} = (\vec{k}_{1,2} \vec{V})$ corresponds to a Doppler shift associated with a non-zero projection of the disk velocity on the direction of light propagation.

Let's combine two thermal fields with different Doppler shifts and mean intensities $\left< I_1 \right>, \left< I_2 \right>$ in an ideal beam splitter (BS). This means that the absolute values of the complex elements of the transfer matrix $2\times 2$ are expressed as $M_{11} = M_{12} = M_{21} = M_{22} = 1/\sqrt{2}$ and the only MEP equals $\pi$. Using Eq.~\eqref{eq:g_1} and the property of the Gaussian distribution of quasi-thermal fields, that the moments of higher orders can always be expressed through the moments of lower orders, one can obtain (see Supplement) an expression for the correlation function of output intensities:
\begin{equation}\label{eq:g2_ideal_BS}
       g^{(2)}(\tau) = 1+ \exp\left(-\frac{\tau ^2}{2 \sigma^2}\right)\times
       \left[\frac{ \left< I_1 \right>^2+\left< I_2 \right>^2}{\left(\left< I_1 \right>+\left< I_2 \right>\right)^2}-\frac{2 \left< I_1 \right> \left< I_2 \right> }{\left[ \left< I_1 \right>+\left< I_2 \right>\right] ^2}\cos \left(d \tau \right)\right].
\end{equation}
Here one can see, that the well-known Gaussian correlation function is modulated with cosine oscillations with the frequency $d\equiv\Delta_2-\Delta_1$ which is equal to the difference between two Doppler shifts. Also one can find the minimum
at $\tau=0$. It means suppression of correlations between two thermal fields at the zero delay as well as it occurs in original HOM interference for two single-photon quantum states. The similar $g^{(2)}$ dependency has been previously observed in~\cite{Chekhova1996, Liu2013}, where HOM effect with thermal fields has been demonstrated in spatial degrees of freedom.

\begin{figure}[htbp]
\centering\includegraphics[width=0.8\linewidth]{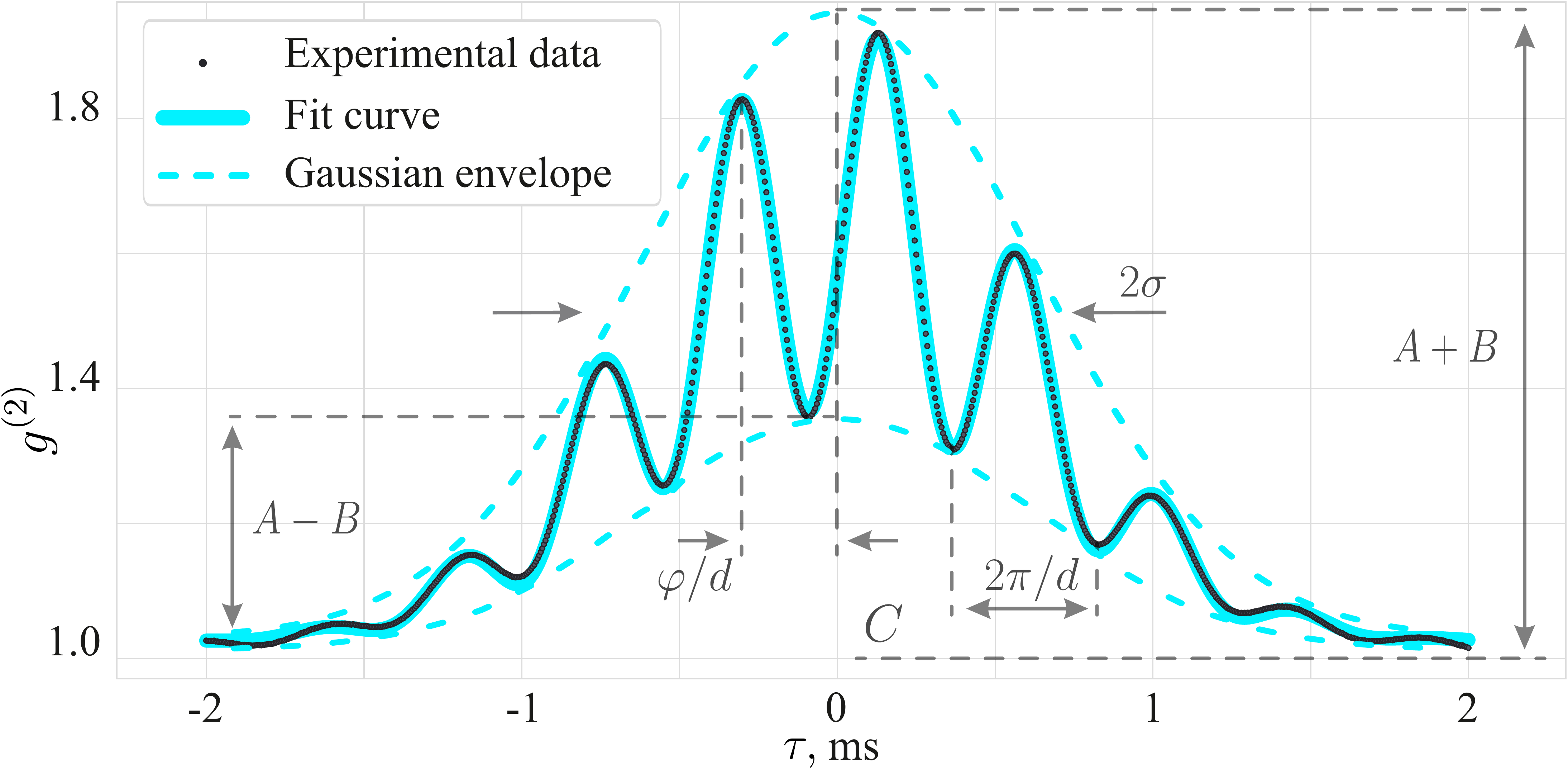}
\caption{Intensity correlation function $g^{(2)}(\tau)$ of two initially uncorrelated thermal fields, interfering at a non-unitary beam splitter (BS) with MEP~$\approx 240^\circ$. Points corresponds to experimental data, solid curve corresponds to the approximating function~\eqref{eq:g2_approx}. 
}
\label{fig:thermal_data}
\label{fig:parameters}
\end{figure}

Let's consider more common case of non-unitary two mode process, where correlation function equals
\begin{equation}\label{eq:g2_approx}
    g^{(2)}\left( \tau  \right) = C + \exp \left( { - {\frac{\tau ^2}{2\sigma^2}}} \right)\left[ {A + B\cos \left( {d \tau  + \varphi_{\text{thermal}} } \right)} \right],
\end{equation}
which plot is shown in Fig.~\ref{fig:parameters}. 
Parameters $A, B, C$ have analytical expressions through absolute values of matrix elements $M_{11}, M_{12}, M_{21}, M_{22}$ and mean values of the input intensities $\left< I_1 \right>$ and $\left< I_2 \right>$ (see details in Supplementary), and the oscillation phase $\varphi_{\text{thermal}}$ exactly equals the desired MEP.

Therefore we propose the following method for characterizing optical circuits. Two uncorrelated thermal fields are injected to the inputs 1 and j of the circuit, the second-order correlation function measured at the outputs 1 and i, and the desired MEP $\varphi_{ij}$ is estimated from the approximation of $g^{(2)}(\tau)$ by Eq.~\eqref{eq:g2_approx}.

Previously, we did a numerical experiment to reconstruct the circuit parameters using the value of $g^{(2)}(\tau =0)$~\cite{Katamadze2021}. In this paper we improve this method, taking into account the time dependence of the correlation function.

The following experiment was carried out to test the proposed method. A bulk optical system $2 \times 2$ with adjustable MEP was designed. Next, some MEP was established and independently determined by two methods: the one proposed above, based on correlation measurements of thermal fields, and the reference one, based on interferometry of coherent states~\cite{Rahimi-Keshari2013, Heilmann2015}. In this way we reconstructed seven different MEPs, and we reconstructed the eighth MEP by taking a $2\times 2$ integrated linear optical circuit.

\begin{figure}[htbp]
\centering\includegraphics[width=0.7\linewidth]{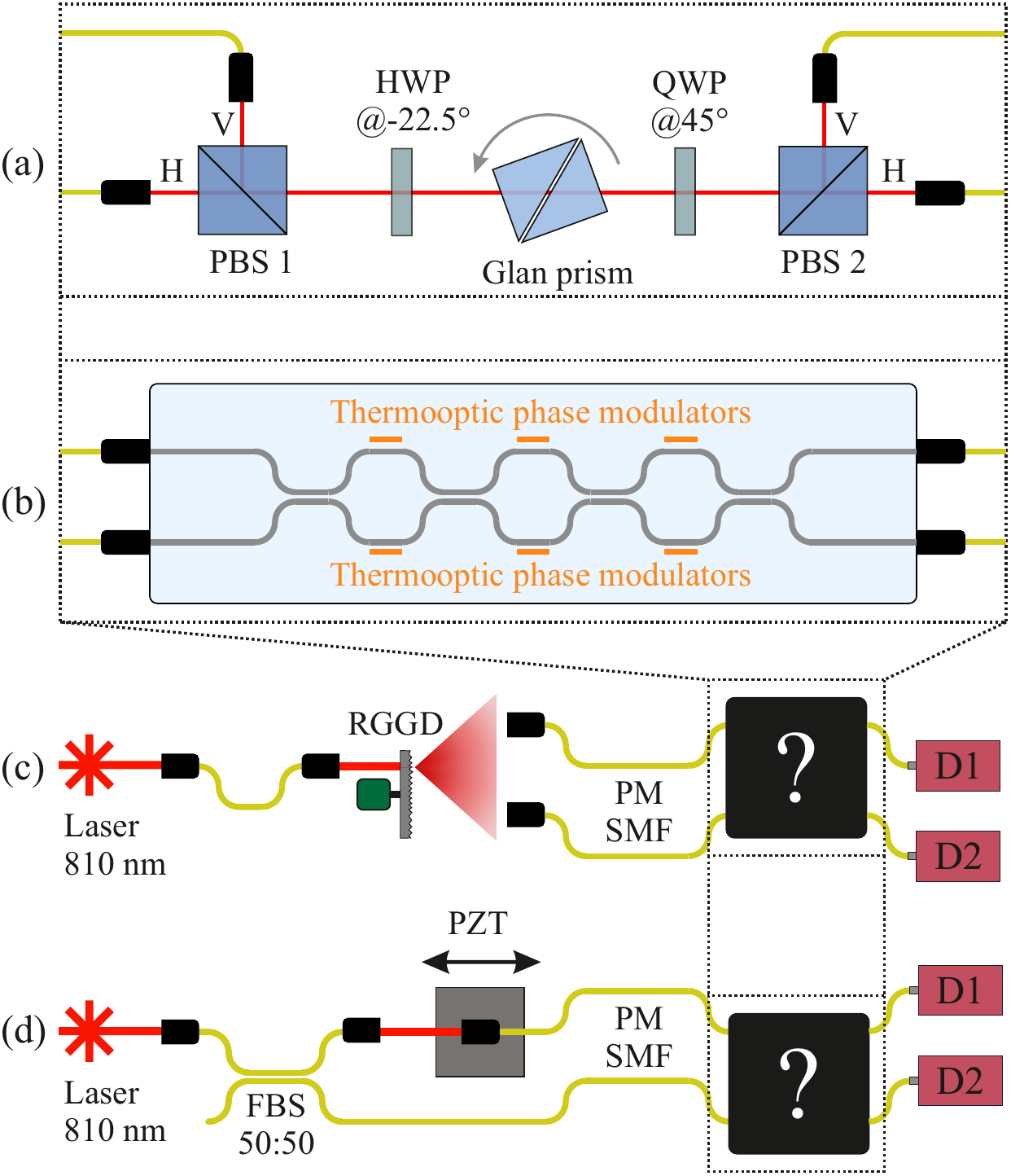}
\caption{General scheme of experimental setup of the black box MEP measurement: (a) non-unitary $2 \times 2$ bulk interferometer where MEP can be adjusted by tilting Glan-Thompson prism, (b) integrated linear optical circuit, (c) setup for characterization of the circuit using two uncorrelated thermal fields, obtained by scattering laser radiation on the rotating ground glass disk (RGGD) and coupled with polarization-maintained single-mode fibers (PM SMF), (d) setup for characterization of the circuit using two coherent states, obtained by laser radiation separated with a symmetric fiber beam splitter (FBS). Their relative phase shift is controlled by the piezotranslator (PZT).
}\label{fig:device}
\end{figure}

A source of uncorrelated quasi-thermal fields with different Doppler shifts is presented in Fig.~\ref{fig:device}c. The radiation of a fiber-pigtailed 810~nm DBR laser (DBR808PN, Thorlabs) was focused on the RGGD and scattered. Then two uncorrelated quasi-thermal fields, propagated at different angles to the disk velocity at the waist point, were coupled with polarization-maintaining single-mode (PM~SMF) fibers by fiber collimators.

For the bulk system to be characterized (Fig.~\ref{fig:device}a), 
we built $2\times2$ interferometer that played the role of a part of some multichannel scheme. 
The input radiation was directed into the circuit through PM SMF and then collimated. The PM fiber axes were oriented orthogonally so the collimated output radiation was combined by a polarizing beam splitter PBS 1 into one beam. Then the radiation passed through the half-wave plate  HWP at -22.5$^\circ$, then PBS~2 again split the radiation into two spatial modes.
Since in the unitary $2 \times 2$ network the MEP is always equal to $\pi$, we introduce adjustable losses by the Glan-Thompson prism, placed between HWP and QWP. By tilting the prism, we could change the transmittion coefficient for horizontal polarization and adjust the MEP in the range of $[-\pi,\pi]$ (see details in Supplement). 

The couplers at the output channels collected radiation into SMF connected to the detectors Thorlabs PDA100A2. Signals were digitized using the ADC device (Rudnev-Shilyaev LA-2USB-14), then the intensity correlation function $g^{(2)} (\tau)$ was calculated from the retrieved data.

For each of the seven setup configurations measurements were carried out 30-40 times and the average MEP and their standard errors were estimated~(see Table~\ref{table1}). Typical experimental values of $g^{(2)}(\tau)$ and their approximation by the Eq.~\eqref{eq:g2_approx} are shown in Fig~\ref{fig:thermal_data}. One can see that the derived function~\eqref{eq:g2_approx} gives an excellent fit of the experimental data. For all the experiments the approximation error ranged from 0.5 to 1.5\%

The eighth MEP value was measured for integrated $2\times 2$ circuit (Fig. \ref{fig:device}b) fabricated using the homebuilt femtosecond laser writing setup 
\cite{Dyakonov2018}. The waveguides were inscribed $15$~$\mu$m below the surface of a fused silica (AG Optics, JGS1 glass) sample using $250$~fs laser pulses emitted at $515$~nm wavelength with $105$~nJ pulse energy. The chip includes a series of Mach-Zehnder interferometers which are equipped with thermo-optic phase modulators. The circuit effectively represents a two-port beamsplitter for any phase configuration selected using the phaseshifters. Losses in the chip are below 1 dB/cm and are homogeneous, so it can be described by a unitary transfer matrix~\cite{Garcia-Patron2019} where the MEP equals $\pi$.

\begin{figure}[ht]
\centering\includegraphics[width=0.7\linewidth]{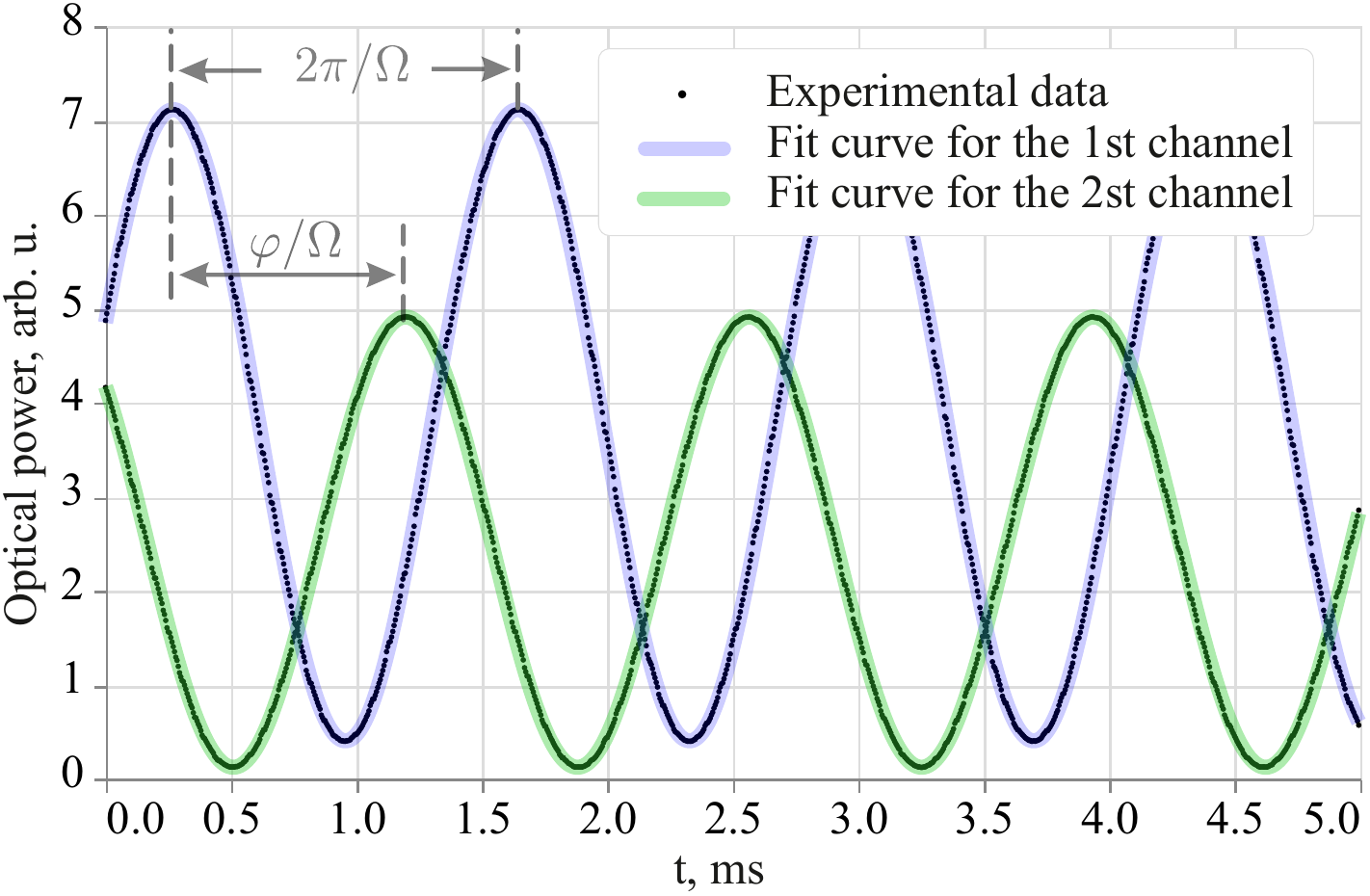}
\caption{Output intensities $I_{1,2}(t)$ of the coherent fields interfering in a non-unitary beam splitter (BS) with MEP~$\approx 240^\circ$, while their relative phase at the inputs of BS is linearly modulated by piezotranslator. Points corresponds to experimental data, solid curve corresponds to the approximating function~\eqref{eq:coher_approx}. Phase difference between these two cosine functions equals the BS MEP.}
\label{fig:coher_data}
\end{figure}

To verify obtained results we used the reference method based on coherent states~\cite{Rahimi-Keshari2013, Heilmann2015}. The laser radiation was separated by a 50:50 fiber beam splitter (FBS). In one of the FBS output channels a controlled phase delay was implemented by a piezotranslator (PZT). Then the radiation was directed into the circuit inputs through the PM~SMF.
To determine the unknown MEP, we approximated the output intensities by cosine functions:
\begin{equation}\label{eq:coher_approx}
    {I_{1,2}}\left( t \right) = {C_{1,2}} + {A_{1,2}}\cos \left[ \Omega (t-t_0) - \phi_{1,2} \right],
\end{equation}
where  $C_{1,2}, A_{1,2}, \Omega$ are unknown parameters that are associated with the average intensities, the visibility and the speed of translator. The parameter $t_0$ is related to the beginning of the time interval where the piezotranslator moved most smoothly. The desired MEP $\varphi_{\text{coherent}}$ is defined as a difference $\phi_1-\phi_2$ (see Fig.~\ref{fig:coher_data}).

\begin{figure}[ht]
\centering\includegraphics[width=0.6\linewidth]{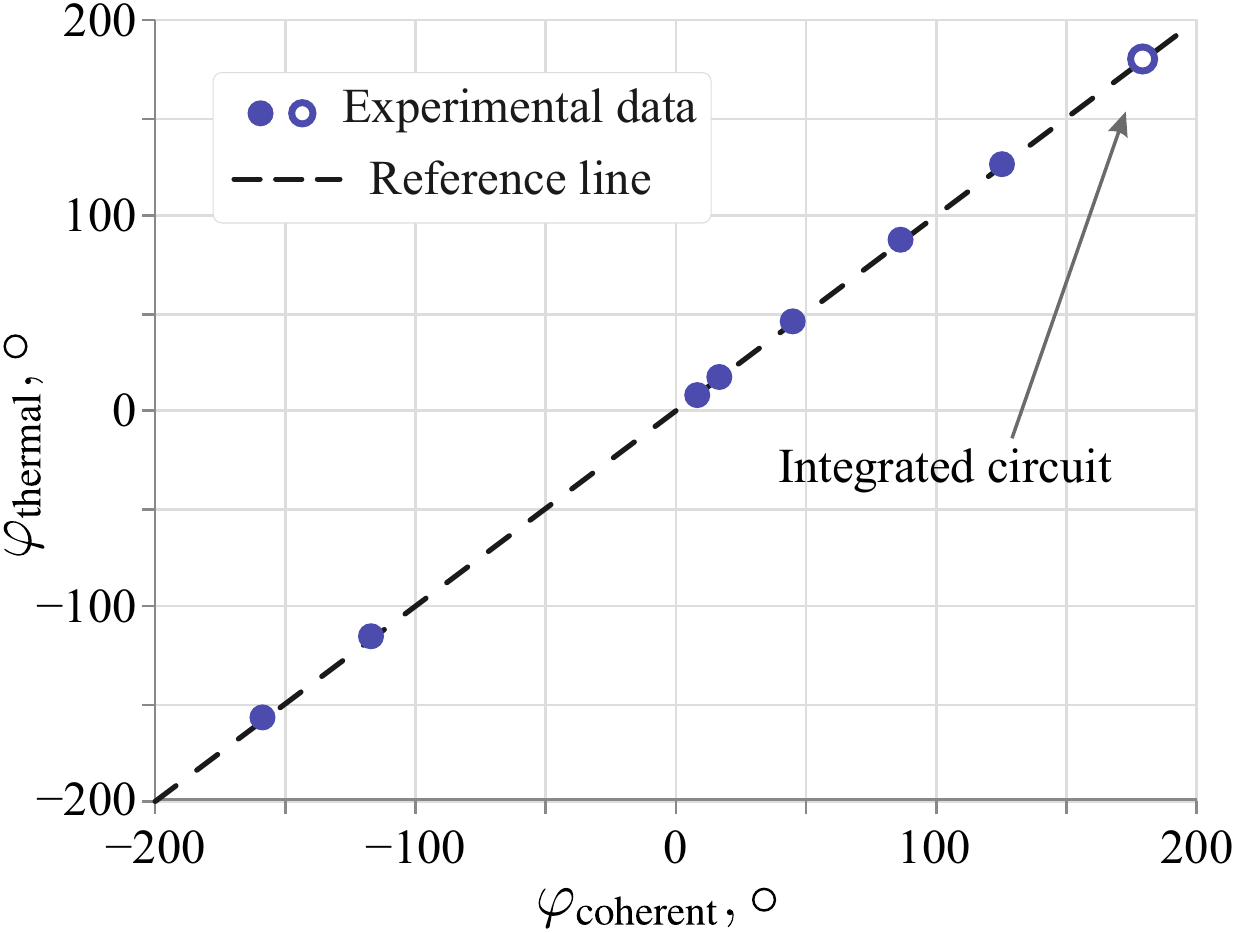}
\caption{The correspondence between the MEPs measured by the thermal fields method ($\varphi_{\text{thermal}}$) and the coherent states method ($\varphi_{\text{coherent}}$). Dashed line corresponds to $\varphi_{\text{thermal
}}=\varphi_{\text{coherent}}$.}
\label{fig:phi}
\end{figure}

The experimental results are shown in Fig.~\ref{fig:phi} and Table~\ref{table1}, where we give average values of MEPs and their standard errors for both methods. For most of the black box configurations the estimated MEP values differ by no more than half of a degree and Fig.~\ref{fig:phi} shows that the results are in a good agreement. However, the Table~\ref{table1} shows that for some experiments the difference between $\varphi_{\text{thermal}}$ and $\varphi_{\text{coherent}}$ is more than $3 \sigma$. This disagreement could be associated with the instability of the black box circuit and time jitter of the acquisition electronics.

\begin{table}[ht]
\caption{\label{table1} MEP values, measured by thermal fields $\varphi_{\text{ thermal}}$ and by coherent fields $\varphi_{\text{ coherent}}$ and their standard errors $\Delta_{\text{thermal}}$  and $\Delta_{\text{coherent}}$ for 8 different black box configurations.}
\vspace{2mm}
\centerline{
\begin{tabular}{ccccc}
\hline
№ & $\varphi_{\text{thermal}}, ^{\circ}$ & $\Delta_{\text{thermal}}, ^{\circ}$ & $\varphi_{\text{coherent}}, ^{\circ}$ & $\Delta_{\text{coherent}}, ^{\circ}$\\
\hline
1& 	-157.650 & 0.113 & -158.274	 & 0.018	 \\
2& -116.101 & 	0.125 & -116.607 & 0.072	  \\
3     & 7.375 & 0.321 & 8.674& 0.006 \\
4  & 16.639 & 0.020 & 17.150 &	0.020 		 \\
5  & 45.130 & 0.125 & 45.331 &	0.037	 \\
6 & 86.919 & 0.233 & 86.751 &	0.036 	 \\
7& 125.597 &	0.314 	&	125.698 &	0.026 \\
8 & 179.463 &	0.180 & 	179.705 &	0.001 	\\
\hline
\end{tabular}}
\end{table}

In summary, we have studied the HOM interference of two originally uncorrelated quasi-thermal fields generated by RGGD. We have found that output cross-correlation function $g^{(2)}(\tau)$ has oscillations, caused by the Doppler effect. These oscillations can be used to characterize linear optical circuits. The MEP reconstruction experiments of the 2-mode optical network have been done. The results are in a good agreement with those obtained by the reference method based on coherent states. The proposed method is robust to the input phase fluctuations and makes it possible to characterize optical circuits effectively, quickly, and inexpensively.

\section*{Funding}
Russian Science Foundation (RSF), project №19-72-10069 (experimental part), Ministry of Science and Higher Education of the Russian Federation, program №FFNN-2022-0016 for the Valiev Institute of Physics and Technology, Russian Academy of Sciences and the Foundation for the Advancement of Theoretical Physics and Mathematics BASIS (project nos. 20-1- 1-34-1, 21-1-3-40-1) (theoretical part).

\section*{Acknowledgments}
This research was performed according to the Development program of the Interdisciplinary Scientific and Educational School of Lomonosov Moscow State University "Photonic and Quantum technologies. Digital medicine". N. N. Skryabin and I. V. Dyakonov acknowledge support from the Russian Roadmap for Quantum computing (Contract No. 868-1.3-15/15-2021 dated October 5, 2021 and Contract No. P2154 dated November 24, 2021)

\section*{Supplemental document}
See Supplement for supporting content.

\bibliographystyle{unsrt} 
\bibliography{biblioteque}

\end{document}


\maketitle

\section{Derivation of the first-order autocorrelation function of  quasi-thermal light}

Let us derive the correlation function of the quasi-thermal field obtained using a disk rotating with an angular velocity $\Omega$ (Fig.~1 in the main text). To describe the field scattered at an angle to the disk surface normal, we follow the paper~\cite{Crosignani1971}. While in that work a plane wave fallen on the disk was bounded by a rectangular window, for our case we consider Gaussian beam, which amplitude is

\begin{equation}
    E(x,y)=E_0 \exp \left(-\frac{(x-L)^2+y^2}{2 w^2}\right).
\end{equation}
Here,  $w$ is a waist of the incident laser beam, which is focusing on the disk surface, $E_0$ corresponds to the field in the center of the beam waist. Also for simplicity we consider the case when the center of the beam waist is at point $\{x=L, y=0, z=0\}$.

Taking into account the intermediate formulas in \cite{Crosignani1971} and equation (13), the dependence of the first-order correlation function on time delay can be written as
\begin{multline}
        g^{(1)}(\tau) \varpropto \exp[\text{i}\omega \tau] \int\limits_{-\infty}^{+\infty}dx''dy''\exp\left[-\text{i}k_y y''(\cos\left(\Omega\tau\right)-1) \right]\exp\left[\text{i}k_y x''\sin\left(\Omega\tau\right)\right]\times\\
    \times E(x'',y''-\Omega\tau L)E(x'', y''),
\end{multline}
where $\omega$ is the frequency of the laser radiation. 
After integrating and using series expansion of the trigonometric functions, one has:

\begin{equation}\label{eq:2_3}
    g^{(1)}(\tau) \varpropto \exp\left[-\cfrac{L^2\Omega^2\tau^2 }{4 w^2}\right]\exp \left[\text{i}(\omega+k_y L \Omega)\tau \right] =\exp\left[-\cfrac{V^2\tau^2 }{4 w^2}\right]\exp \left[\text{i}(\omega+\Delta)\tau \right] .
\end{equation}
Here, $V$ is speed of the disk point at the center of the waist, the first factor is the standard Gaussian envelope corresponding to the correlation properties of the thermal state, with the correlation time $w/V$, which corresponds to the time of passage of the disk point through the beam waist, and the second factor is sinusoidal oscillation, where $\omega$ is the frequency of the radiation incident on the disk, and $\Delta = k_y V$ corresponds to the Doppler shift associated with a non-zero projection of the disk velocity on the direction of light propagation.

\section{Intensity correlation function of two interfering quasi-thermal fields}

The Gaussian (thermal) distribution is remarkable by its higher-order moments can be expressed in terms of first-order moments.

Let's put the intensities at inputs $1$ and $j$ as $I_1$ and $I_j$, and intensities at outputs $1$ and $i$ as $I_1^{\text{(out)}}$ and $I_i^{\text{(out)}}$. In these terms the output second-order cross-correlation function equals:

\begin{multline}\label{eq:124}
    \left\langle {I_1^{\text{\text{(out)}}}I_i^{\text{\text{(out)}}}(\tau )} \right\rangle  = \left\langle {E_1^{\text{(out)}*} E_i^{\text{(out)}*} (\tau ){E_1^{\text{\text{(out)}}}}{E_i^{\text{\text{(out)}}}}(\tau )} \right\rangle  = \\
    = {\left\langle {E_1^{\text{(out)}*} {E_1^{\text{\text{(out)}}}}} \right\rangle } {\left\langle {E_i^{\text{(out)}*} (\tau ){E_i^{\text{\text{(out)}}}}(\tau )} \right\rangle } + \left\langle {E_1^{\text{(out)}*} {E_i^{\text{\text{(out)}}}}(\tau )} \right\rangle \left\langle {E_i^{\text{(out)}*} (\tau ){E_1^{\text{\text{(out)}}}}} \right\rangle \\ = \left\langle I_1^{\text{\text{(out)}}} \right\rangle \left\langle I_i^{\text{\text{(out)}}} \right\rangle +\left\langle {E_1^{\text{(out)}*} {E_i^{\text{\text{(out)}}}}(\tau )} \right\rangle \left\langle {E_i^{\text{(out)}*} (\tau ){E_1^{\text{\text{(out)}}}}} \right\rangle
\end{multline}

We are interested in the last term. Let's describe it in more detail:
\begin{multline}
    \left\langle {E_1^{\text{(out)}*} {E_i^{\text{(out)}}}(\tau )} \right\rangle  = {\left\langle {E_i^{\text{(out)}*} (\tau ){E_1^{\text{(out)}}}} \right\rangle ^ * } = \\ = \left\langle {\left( {M_{11}E_1^*  + M_{1j}E_j^* } \right)\left( {M_{i1}{E_1}(\tau ) + M_{ij}{{\mathop{\rm e}\nolimits} ^{i\varphi }}{E_j}(\tau )} \right)} \right\rangle  = \\
 = M_{11}M_{i1}\underbrace {\left\langle {E_1^* {E_1}(\tau )} \right\rangle }_{\left\langle {I_1} \right\rangle g_1^{(1)}(\tau )} + M_{11}M_{ij}{{\mathop{\rm e}\nolimits} ^{\text{i}\varphi }}\underbrace {\left\langle {E_1^* {E_j}(\tau )} \right\rangle }_0 +\\+ M_{1j}M_{i1}\underbrace {\left\langle {E_j^* {E_1}(\tau )} \right\rangle }_0 +M_{1j}M_{ij}{{\mathop{\rm e}\nolimits} ^{i\varphi }}\underbrace {\left\langle {E_j^* {E_j}(\tau )} \right\rangle }_{\left\langle {I_j} \right\rangle g_j^{(1)}(\tau )},
\end{multline}

\begin{multline}\label{eq:2_4}
    { {\left\langle {E_1^{\text{(out)}*} {E_i^{\text{(out)}}}(\tau )} \right\rangle } ^2} = { {M_{11}M_{i1}\left\langle {I_1} \right\rangle g_1^{(1)}(\tau ) + M_{1j}M_{ij}{{\mathop{\rm e}\nolimits} ^{i\varphi }}\left\langle {I_j} \right\rangle g_j^{(1)}(\tau )}^2} = \\
 \left\langle {I_1} \right\rangle \left\langle {I_j} \right\rangle M_{11}  M_{1j}  M_{ij} M_i1e^{-\text{i} \varphi } g_1^{(1)}(\tau ) g_j^{(1)}(\tau )^*+
 \left\langle {I_1} \right\rangle \left\langle {I_j} \right\rangle M_{11}  M_{1j}  M_{ij} M_i1 e^{\text{i} \varphi } g_j^{(1)}(\tau ) g_1^{(1)}(\tau )^*+\\
 \left\langle {I_1} \right\rangle^2 M_{11}^2 M_{i1}^2 g_1^{(1)}(\tau ) g_1^{(1)*}(\tau )+\left\langle {I_j} \right\rangle^2 M_{1j} ^2 M_{ij}^2 g_j^{(1)}(\tau ) g_j^{(1)*}(\tau )
 .
\end{multline}
 	
Now we replace the first-order correlation functions here with \eqref{eq:2_3}. Let them differ in Doppler shift for fields at different inputs. This can be achieved by collecting radiation scattered on the same RGGD at different angles. Then assuming
 	
\begin{gather}
    g_1^{(1)}(\tau ) = \exp \left(  - \frac{V^2\tau ^2}{4w^2} \right)\exp \left[ {  \text{i}\left( {{\omega} + {\Delta _1}} \right)\tau } \right],  \\
    g_j^{(1)}(\tau ) = \exp \left(  - \frac{V^2\tau ^2}{4w^2} \right)\exp \left[ {  \text{i}\left( {{\omega} + {\Delta _j}} \right)\tau } \right],\\
\text{and } d = {\Delta _j} - {\Delta _1}
\end{gather}
for the normalized cross-correlation function we have:
%
\begin{multline}\label{eq:g2t}
    g^{(2)}(\tau) = 1 + \exp{\left(-\frac{\tau ^2 V^2}{2 w^2}\right)} \frac{1}{\left(\left\langle I_1 \right\rangle M_{11}^2+\left\langle I_j \right\rangle M_{1j}^2\right) \left(\left\langle I_1 \right\rangle M_{i1}^2+\left\langle I_j \right\rangle M_{ij}^2\right)} \times \\
    \left[\left\langle I_1 \right\rangle^2 M_{11}^2 M_{i1}^2 +\left\langle I_j \right\rangle^2 M_{1j}^2 M_{ij}^2 +2 \left\langle I_1 \right\rangle \left\langle I_j \right\rangle M_{11} M_{i1} M_{1j} M_{ij} \cos \left(d \tau +\varphi \right)\right] .
\end{multline}
In the special case of the ideal symmetric beam splitter ($M_{11} = M_{12} = M_{21} = M_{22} = 1/\sqrt{2}$ and the only MEP equals $\pi$) \eqref{eq:g2t} is reduced to Eq.~(4) of the main article:
\begin{equation*}\label{eq:g2_ideal_BS}
       g^{(2)}(\tau) = 1+ \exp\left(-\frac{\tau ^2}{2 \sigma^2}\right)\times
       \left[\frac{ \left< I_1 \right>^2+\left< I_2 \right>^2}{\left(\left< I_1 \right>+\left< I_2 \right>\right)^2}-\frac{2 \left< I_1 \right> \left< I_2 \right> }{\left[ \left< I_1 \right>+\left< I_2 \right>\right] ^2}\cos \left(d \tau \right)\right].
\end{equation*}

In general case, \eqref{eq:g2t} can be represented as Eq.~(5) from the main article:
\begin{equation*}\label{eq:g2_approx}
    g^{(2)}\left( \tau  \right) = C + \exp \left( { - {\frac{\tau ^2}{2\sigma^2}}} \right)\left[ {A + B\cos \left( {d \tau  + \varphi_{\text{thermal}} } \right)} \right],
\end{equation*}

where parameters:

\begin{equation}
\begin{aligned}
A &= \frac{\left\langle I_1 \right\rangle^2 M_{11}^2 M_{i1}^2 +\left\langle I_j \right\rangle^2 M_{1j}^2 M_{ij}^2}{\left(\left\langle I_1 \right\rangle M_{11}^2+\left\langle I_j \right\rangle M_{1j}^2\right) \left(\left\langle I_1 \right\rangle M_{i1}^2+\left\langle I_j \right\rangle M_{ij}^2\right)};\\
B &= \frac{2 \left\langle I_1 \right\rangle \left\langle I_j \right\rangle M_{11} M_{1j} M_{ij} M_{i1}}{\left(\left\langle I_1 \right\rangle M_{11}^2+\left\langle I_j \right\rangle M_{1j}^2\right) \left(\left\langle I_1 \right\rangle M_{i1}^2+\left\langle I_j \right\rangle M_{ij}^2\right)};\\
C &=1;\\
\sigma &= \frac{w}{V}.
\end{aligned}
\end{equation}

\section{Implementation of the beam splitter with adjustable matrix element phase}

\begin{figure}[htbp]
\centering\includegraphics[width=0.7\linewidth]{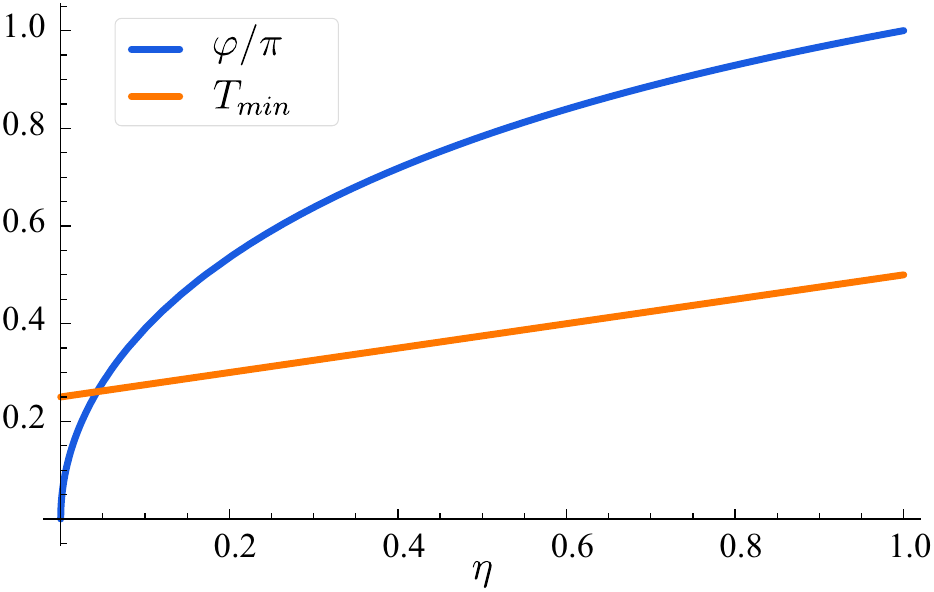}
\caption{MEP value $\varphi/\pi$ and minimum intensity transmission coefficient  $T_{min}$ vs Glan prism transmission coefficient $\eta$ for horizontal polarization, which can be tuned by rotation of the prism. }\label{fig:phase_eta}
\end{figure}

The arrangement of the polarization elements (Fig.~2 in the main text) in the order PBS, HWP at -22.5$^\circ$, tilted Glan-Thompson prism, and quarter-wave plate QWP at 45$^\circ$ and second PBS allows us to change the MEP of the transfer matrix from $0$ to $\pi$. Swapping the outputs of the circuit, we get the same matrix element phase (MEP) with the opposite sign. Thus, we can change the MEP from $-\pi$ to $\pi$. The  Fig.~\ref{fig:phase_eta} shows the dependence of the MEP and minimum intensity transmission coefficient $T_{min}$ on the prism transmission coefficient of horizontally polarized beam $\eta$. It can be seen that the MEP varies from $0$ to $\pi$, without zeroing out the transmission anywhere.

When calculating this dependence, we used Jones calculus:
\begin{equation}
   \vec{H}=\left(
\begin{array}{c}
 1 \\
 0 \\
\end{array}
\right),
\Vec{V}=\left(
\begin{array}{c}
 0 \\
 1 \\
\end{array}
\right)
\end{equation}

Glan-Thompson prism matrix can be represented as

\begin{equation}
    \hat{P}=\left(
\begin{array}{cc}
 \sqrt{\eta} & 0\\ 
 0 & 1
\end{array}
\right),
\end{equation}
where $\eta$ is transmission coefficient for muted polarization, which can be changed by tilting the prism around the vertical axis. In our case, in the initial position of the prism, when $\eta=0$, the horizontal polarization is not transmitted.

Matrices of wave plates are represented as
\begin{equation}
    \hat{HWP} = \left(
\begin{array}{cc}
 \cos ^2(\alpha)-\sin ^2(\alpha) & 2 \sin (\alpha) \cos (\alpha) \\
 2 \sin (\alpha) \cos (\alpha) & \sin ^2(\alpha)-\cos ^2(\alpha) \\
\end{array}
\right),
\end{equation}

\begin{equation}
    \hat{QWP} = \left(
\begin{array}{cc}
 \cos ^2(\beta)+\text{i} \sin ^2(\beta) & (1-\text{i}) \sin (\beta) \cos (\beta) \\
 (1-\text{i}) \sin (\beta) \cos (\beta) & \sin ^2(\beta)+\text{i} \cos ^2(\beta) \\
\end{array}
\right),
\end{equation}
where $\alpha$ and $\beta$ are angles of fast axes.

The resulting matrix $\hat{U}$ will be

\begin{equation}
    \hat{U} = \hat{HWP} \times \hat{P} \times \hat{QWP}.
\end{equation}
Then we can find an equivalent matrix with zeroed first row and column and determine the MEP of the element $\{2,2\}$ and the minimal transmission coefficient $T_{min}$.

\bibliographystyle{unsrt} 
\bibliography{biblioteque}